\documentclass[preprint,superscriptaddress]{revtex4-1}%
\usepackage{graphicx}

\usepackage{units}
\usepackage{amsmath}
\usepackage{textcomp}

\usepackage{stackengine}
\stackMath
\let\svoverline\overline
\def\overline#1{\stackengine{0pt}{}{\svoverline{\phantom{#1}}}{O}{l}{F}{T}{L}#1%
}

\begin{document}


\title{Intervalley scattering in MoS$_2$ imaged by two-photon photoemission
with a high-harmonic probe} 



\author{R. Wallauer}
\author{J. Reimann}
\author{N. Armbrust}
\author{J. G\"udde}
\author{U. H\"ofer}
\affiliation{Fachbereich Physik und Zentrum f{\"u}r
Materialwissenschaften, Philipps-Universit{\"a}t, 35032 Marburg,
Germany}


\date{\today}

\begin{abstract}
We report on the direct mapping of electron transfer in the
momentum space of bulk MoS$_2$ by means of time- and
angle-resolved two-photon photoemission with a high-harmonic
probe. For this purpose, we have combined a high-repetition rate
high-harmonic source with tunable femtosecond pump pulses and a 3D
($k_x, k_y, E$) electron spectrometer.
 We show that optical excitation slightly above the A exciton
resonance results in an immediate occupation of the conduction
band at $\overline{K}$ followed by an ultrafast transfer ($<
50$~fs) to the conduction band minimum at $\overline{\Sigma}$.
 Both signals, at $\overline{K}$ and $\overline{\Sigma}$, do not
vanish over the observed period of 400~fs. The technique described
 here enables direct access to the charge transfer dynamics in
$k$-space and allows the study of decay times and decay channels
in various systems with dependence on the excess energy or
helicity of the excitation.
\end{abstract}


\maketitle 



Transition-Metal Dichalchogenides (TMDC's) have been widely studied
because of their rich phase diagram that includes superconductivity,
charge-density waves or metal insulator transitions.
 With the recent emergence of TMDC atomically thin films
\cite{Mak10prl, Splend10nl}, the interest in these materials has
increased dramatically also due to potential applications, like
switching and optoelectronic devices \cite{Radisa13natmat,
Yin12acsnano}, energy storage \cite{Ding12nanoscale} or nanodevices
\cite{Wu14nature}. Heterostructures of MoS$_2$ and graphene or other
TMDC materials add another degree of freedom to tailor material
properties and are an important step toward functional devices
\cite{Shi12nl, Duan14natnano, Geim13nature}.
 Despite the tremendous efforts in this field, there still remains strong
disagreement on fundamental properties.
 Measurements of carrier mobilities, for example, differ by one order
of magnitude \cite{Radisa13natmat, Strait14prb} and decay times
extracted from time-resolved absorption measurements even differ by
up to two orders of magnitude \cite{Seo16scirep, Manneb14acsnano,
Shi13acsnano}.
 In the case of monolayer films, many of these discrepancies may
be related to the influence of the substrate or sample quality.
 However, even for bulk or few-layer samples, electron relaxation
processes are not fully understood and only a small number of
time-resolved experiments have been reported \cite{Tanaka03prb,
Kumar13jap}.
 A common problem of many experiments is the indirect nature of the
measurements.
 In pure optical experiments, for example, many possible excitation
and relaxation pathways can contribute to the measured transients
which makes the identification of transfer and decay channels
difficult.

 For single-crystalline samples, time- and angle-resolved two-photon
photoemission (2PPE) can provide this information in many cases as
it is capable of directly imaging the electron dynamics in k-space
\cite{Haight95ssr, Rohled05njp}. Electronic states at surfaces and
in two-dimensional or layered materials are delocalized in two
directions and well localized in the perpendicular direction. The
excited electrons are then characterized by their energy and
parallel momentum, i.e. quantities that 2PPE accesses directly by
measuring the kinetic energy $E_{kin}$ and emission angle $\theta$
of the photoelectrons, just like standard angle-resolved
photoemission (ARPES) of occupied states. In contrast to ARPES,
angle-resolved 2PPE is often limited to the detection of small
parallel momenta ($\lesssim 0.5$~\AA$^{-1}$) around
$\overline{\Gamma}$ since it typically employs photon energies only
up to 6.5 eV which can  be generated in available nonlinear crystals
\cite{Ge98sci,Bertho02prl,Schmitt08sci}. It can be seen easily from
the relationship $k_{\parallel} (\AA^{-1}) = 0.514\, \sqrt{
E_{kin}(eV)} \sin\theta$ that photon energies of 20 eV and higher
are required in order to probe the full Brillouin zone of typical
materials.

High harmonic generation (HHG) in noble gas jets makes it possible
to produce ultrashort pulses in this energy range and utilize them
for time-resolved photoemission experiments \cite{Haight95ssr,
Bauer01prl}.
 First angle-resolved 2PPE experiments with high-harmonic probe
studied collective phenomena after strong excitations in the
occupied band structures such as phase transitions in charge-density
wave materials and magnetic systems \cite{Rohwer11nat, Peters11prl,
Fritsc13rsi} or hot electron dynamics close to the Fermi level $E_F$
as in graphene \cite{Gierz15prl, Johann13prl}.
 In these experiments, the specific photon energy of the excitation
played a minor role.
 Only recently, an example has been shown where, for a sub-monolayer
of MoS$_2$ on an Au(111) substrate, tunable pump pulses were used to
drive a specific transition into unoccupied bands far above $E_F$
\cite{Cabo15nl}.
 This is also the only study on a MoS$_2$ based
system in which the ultrafast excitation and decay at the
$\overline{K}$ point have been directly observed.
 All of the aforementioned experiments were based on laser systems
with high pulse energies at repetition rates in the few kHz regime
which make it easy to reach the high intensities that are required
for HHG.
 The signal-to-noise ratio in photoemission spectroscopy, however,
greatly benefits from higher repetition rates which also reduce
space charge effects that are not only relevant for the
high-harmonic probe pulses \cite{Passla06jap}, but also for the pump
pulses \cite{Oloff16jap}.
 In addition, the applied pump fluence has to be limited if we want
to study single-particle dynamics rather than collective processes,
making a higher repetition rate even more desirable.

\begin{figure}[t]
    \includegraphics[scale=0.17]{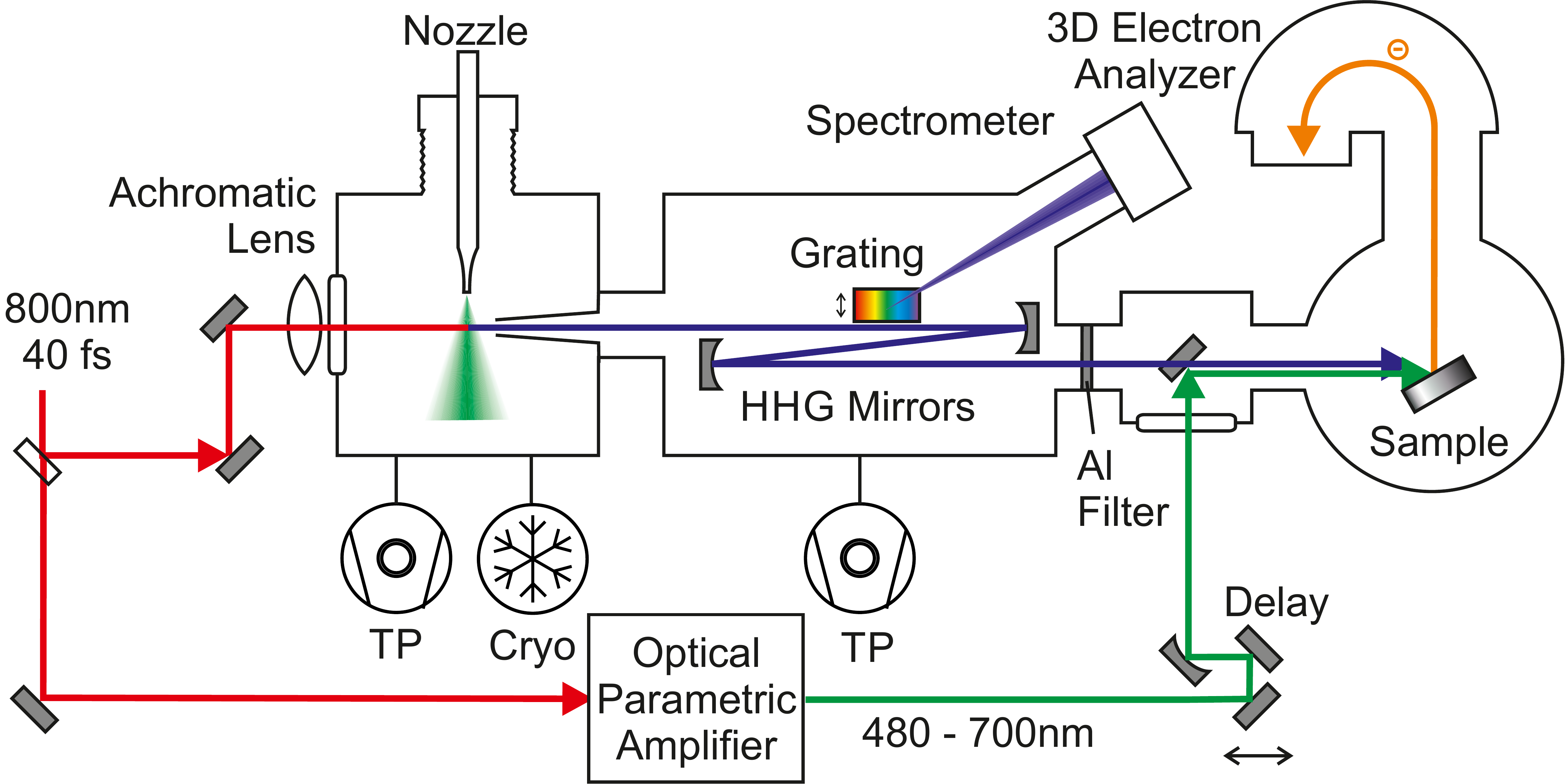}
    \caption{\label{fig:setup}
    Sketch of the experimental setup. Pump pulses are frequency
    converted by an optical parametrical amplifier. Probe pulses are
    focused into a Xe-jet and the generated harmonics are selected by
    two multi-layer mirrors. Both pulses are overlapped before entering
    the analyzer chamber equipped with a Scienta DA30 hemispherical
    analyzer.}%
\end{figure}

In this letter, we demonstrate that 2PPE experiments with HHG probe
pulses are feasible at high repetition rates and charge transfer
after excitation to the conduction band with visible pump pulses can
be directly mapped in $k$-space.
 We chose bulk MoS$_2$ as a prototypical example for the observation
 of charge transfer dynamics in $k$-space.
 For this material, a direct transition at $\overline{K}$ is
accessible with our pump pulses and the transfer to the conduction
band minimum at the $\overline{\Sigma}$ point is suggested by many
experimental and theoretical studies.
 The experiments were carried out on a setup which has been developed
in our group \cite{Heyl12jpb}.
 A sketch is shown in Fig. \ref{fig:setup}.
 It combines a high-repetition rate laser system for high-harmonic
generation, tunable pump pulses in the visible regime, and a 3D
electron spectrometer which is capable of energy and
k$_x$,k$_y$-momentum mapping in both directions parallel to the
surface.
 The output of a 800~nm, 50~fs, 100~kHz, 1~W regenerative amplifier
is split 50/50 into one pump and one probe arm.
 Pump pulses are frequency converted by an optical parametric
amplifier, which is tunable between 480 and 700~nm, with a typical
pulse energy of 200~nJ.
 The pump beam is focused outside the UHV chamber by an $f=500$~mm
spherical mirror and redirected inside the UHV by a cutted mirror to
enter the analyzer chamber almost collinear with the HHG-beam.
 With a spot size well below 200~$\mu$m, we achieved pump fluences up
to 1~mJ/cm$^2$.

The generation of high harmonics in this setup is explained in great
detail in \cite{Heyl12jpb}.
 Here, we summarize the most important points for this experiment and
significant improvements.
 The laser pulses of the probe arm are focused by an achromatic
$f=60$~mm lens into a Xenon gas jet in an ultrahigh vacuum chamber
generating high harmonics up to the $21^{\rm st}$ harmonic of the
fundamental.
 The Xenon gas jet is generated by a glass nozzle with an inner
diameter of 30~$\mu$m at a backing pressure of 4~bar.
 An additional cryopump not only increases pumping speed in the jet
chamber, but also enables us to efficiently recycle the Xenon.
 While operating the jet, we connect the output of the turbo pump to
the cryopump and are thereby able to freeze all of the used Xenon
within the cryopump.
 After continuous operation over several days, the Xenon is then defrosted
into a 50~l stainless steel vessel and liquified in a half liter gas
cylinder cooled by liquid nitrogen. From this procedure, we are able
to recycle almost all of the used Xenon.

\begin{figure}[t]
    \includegraphics[scale=0.45]{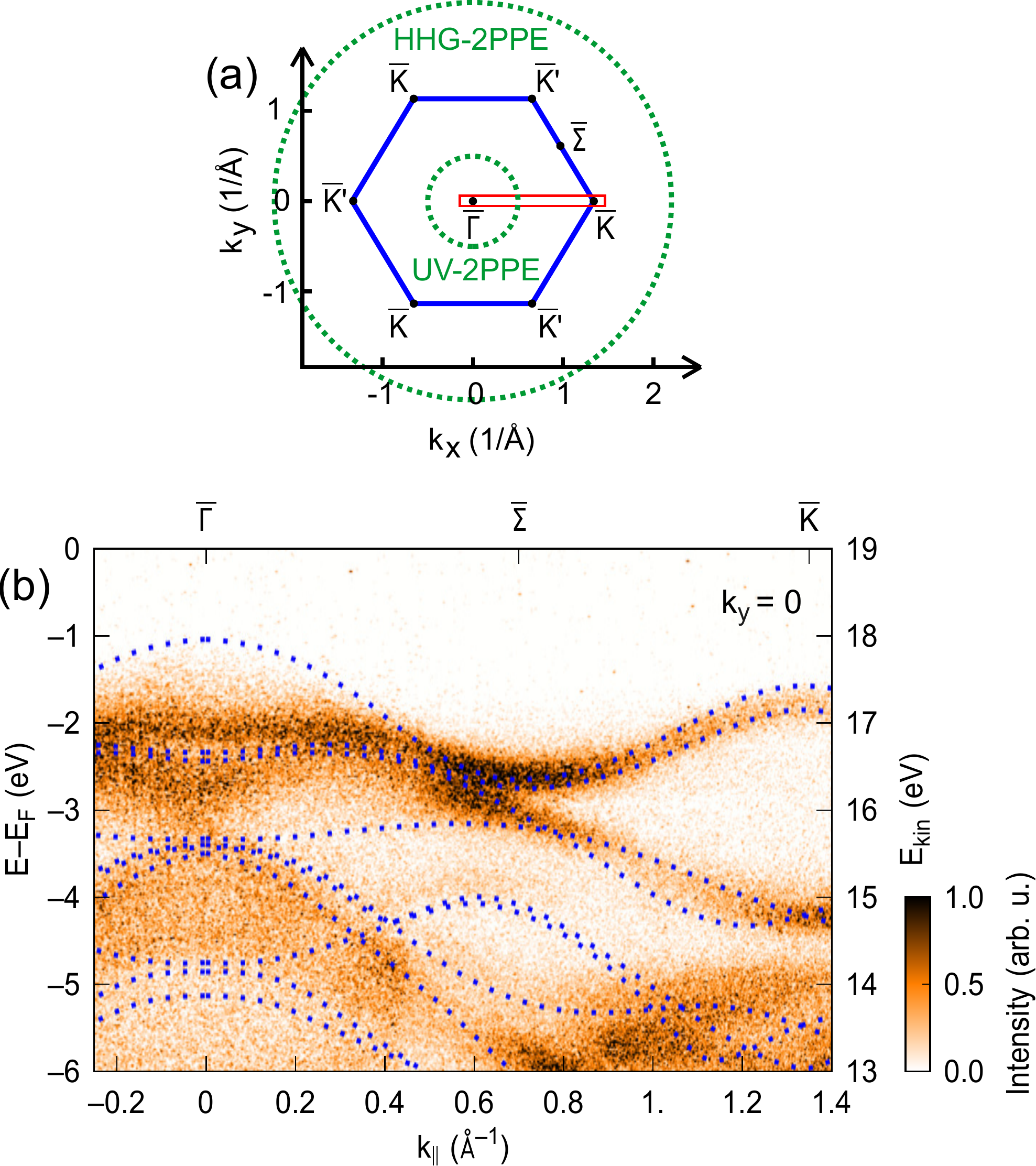}
    \caption{\label{fig:brillouin_zone}
    (a) Brillouin zone of MoS$_2$ with important high symmetry points.
    Measurements were carried out along
    ${\protect\overline{\Gamma}} - {\protect\overline{K}}$ as indicated
    by the red bar. Dashed green lines illustrate the increase in
    observable parallel momentum when using probe pulses generated
    by high harmonics (HHG-2PPE) in contrast to typical UV pulses (UV-2PPE).
    (b) Measured energy-momentum map
    in the ${\protect\overline{\Gamma}} - {\protect\overline{K}}$
    direction. Blue dashed lines are theoretical
    band structure calculations.}
 \end{figure}

In order to characterize and optimize the harmonic spectrum, a
grating (600~lines/mm) can be moved into the beam and the
diffraction pattern is detected on a multichannel plate with a
phosphorus screen.
 The calibration of the HHG intensity was done by extrapolating
single-photon counts at low intensities taking into account the
efficiency of the grating and the MCP as well as the transmission
through the filter for an absolute value of the number of photons
per second.
 By varying the distance of the nozzle from the focus, both along and
perpendicular to the optical axis, as well as the gas pressure, we
are able to optimize the flux of a specific harmonic.
 We achieved the highest overall flux of $> 10^{10}$~photons/s
for the $15^{\rm{th}}$ harmonic (23.25~eV). In order to select a
single harmonic, we use a pair of multi-layer mirrors (IOF Jena)
with a reflectivity of larger than 30\% while suppressing about 99\%
of neighboring harmonics.
 The accessible momentum space for this photon energy is illustrated
in Fig. \ref{fig:brillouin_zone}a where the Brillouin zone of
MoS$_2$ is drawn in blue. Green dashed circles indicate the
photoemission horizon for a typical 2PPE using UV probe pulses
(inner circle) and HHG at 23.25~eV (outer circle).

Commercial 2H-MoS$_2$ single crystals (2D Semiconductors) were
cleaved {\it in situ} under ultra high vacuum conditions at a
pressure of better than $10^{-10}$~mbar.
 The orientation and surface quality was checked by means of LEED.
 We oriented all samples with the analyzer slit parallel to the
$\overline{\Gamma} - \overline{K}$ direction as indicated by the red
rectangle in Fig. \ref{fig:brillouin_zone}a in order to map the
$\overline{\Sigma}$ and the $\overline{K}$-point within the same
measurement at $k_y = 0$.
 All spectra are recorded at room temperature with a Scienta DA30
hemispherical analyzer.
 Fig. \ref{fig:brillouin_zone}b shows the recorded two-dimensional
energy versus momentum maps in this direction, excited solely with
the HHG pulses.
 The measured band structure is in very good agreement with
theoretical calculations, taken from \cite{Latzke15prb} (blue dashed
lines).
 The absence of spectral weight in the valence band at the
$\overline{\Gamma}$-point has been observed in ARPES measurements
before \cite{Boeker01prb, Latzke15prb} and is attributed to a matrix
element effect.
 We observe the spin splitting of the valence band at $\overline{K}$ when choosing
a smaller entrance slit ($0.5$~mm) of the hemispherical analyzer
(not shown).
 Since the time-resolved measurements are more limited by count rate
than by energy resolution, we usually measure at large slit settings
($2.5$~mm).

\begin{figure*}[t]
    \includegraphics[scale = 0.45]{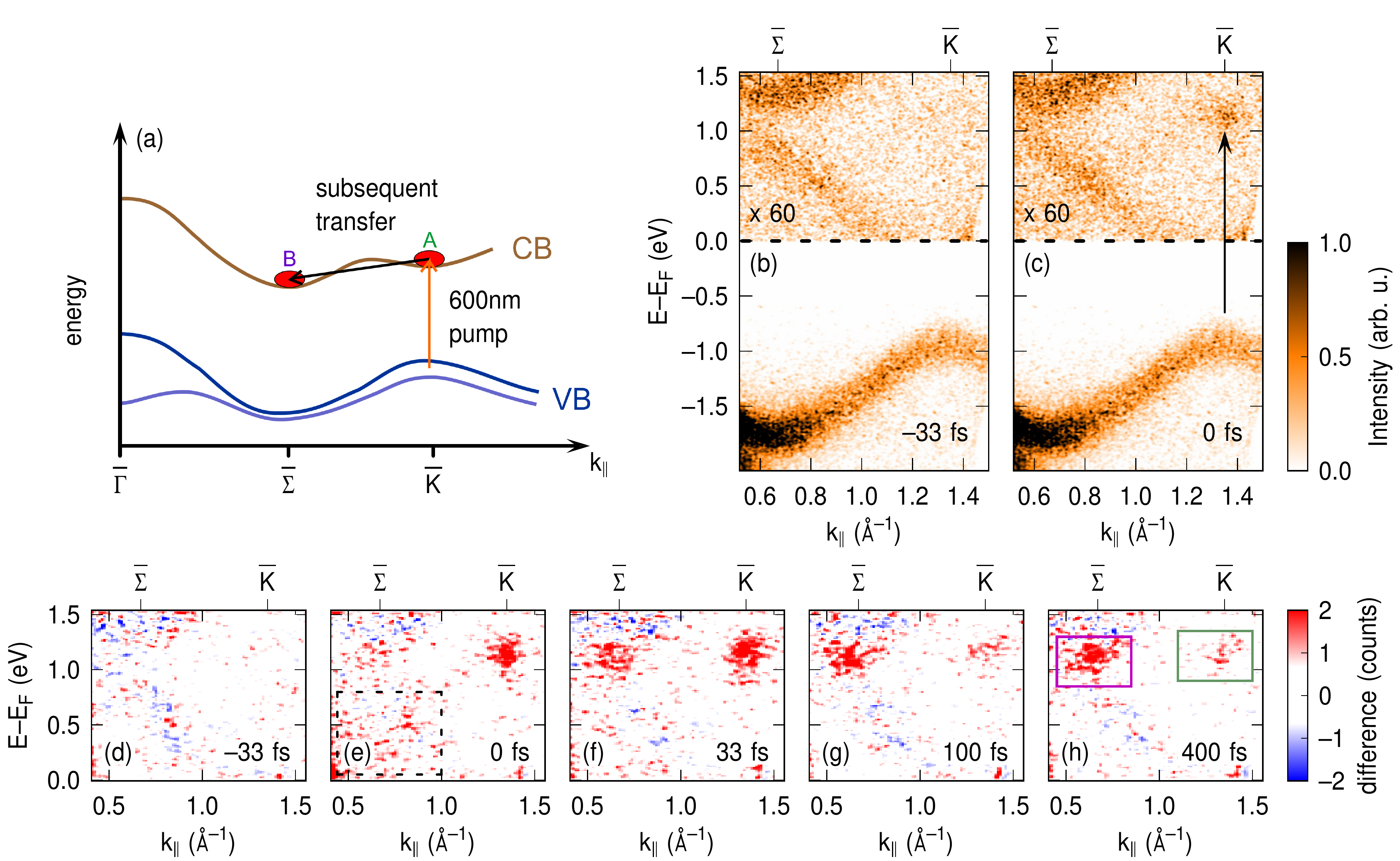}
    \caption{\label{fig:excitation}
    (a) Excitation scheme of the experiment. Electrons are excited at
    $\protect\overline{K}$ by pump pulses of 600~nm slightly above the A exciton
    resonance. They are subsequently transferred to the conduction
    band minimum at $\protect\overline{\Sigma}$.
    (b) and (c) show energy-momentum maps before and at
    temporal overlap of the two laser pulses, respectively. Each map is
    composed of measurements of the occupied in the lower part and
    the unoccupied part in the upper part, separated by the Fermi energy
    (dashed black line). The intensity of the unoccupied part is amplified by a
    factor of $\sim 60$. The transiently populated conduction band in the latter case
    is clearly visible. (d)-(h) Difference spectra at various time
    delays for illustration of the charge transfer. Even for the longest
    delay (h) the population at $\protect\overline{K}$
    remains visible. Additional intensity around temporal overlap in the
    lower left corner is assigned to non-resonant excitation.}
\end{figure*}

Fig. \ref{fig:excitation}a shows a sketch of the excitation process
used for the time-resolved measurements.
 We pump with 600~nm ($2.05$~eV) photons, which is at least 200~meV
above the A exciton resonance excitation.
 While the exciton binding energy in monolayer MoS$_2$ is predicted
by theory to be up to 1~eV \cite{Qiu13prl}, experiments on bulk
sample report a value of less than 100~meV \cite{Saigal16apl}.
 We therefore expect for the optical excitation an interband
transition from the valence to the conduction band close to the
$\overline{K}$-point.
 After excitation, the electrons can subsequently decay to
$\overline{\Sigma}$ where the global conduction band minimum is
predicted by all band structure calculations.
 Our time-resolved experiments show this transfer directly.
 Only recently, this transfer has also been observed in the very
similar material WSe$_2$ \cite{Berton16arxiv}.

 While for the band structure map shown in
Fig.~\ref{fig:brillouin_zone}b three spectra were stitched together
in order to cover the complete momentum path from
$\overline{\Gamma}$ to $\overline{K}$, for the time-resolved
measurements we chose a sample orientation where we can measure a
single energy map that includes both $\overline{K}$ and
$\overline{\Sigma}$.
 In Fig. \ref{fig:excitation} we show two of such energy maps at
different delay times, before the excitation (negative delay) (b)
and at temporal overlap (c).
 Both spectra were combined from measurements of the occupied states
and unoccupied states, while the unoccupied part has been
intensified by a factor of $\sim 60$ for better visibility.
 Each spectrum was measured for 30~min.
 Clearly, the population of the conduction band at the
$\overline{K}$-point can be observed at temporal overlap even in the
raw data.
 The part of the intensity distribution in the unoccupied part,
which is also visible at negative delays, can be assigned to the
residual neighboring $17^{\rm{th}}$ harmonic.
 In this case, replica of the occupied band structure are shifted
by $3.1$~eV and superimpose the unoccupied part.
 This particularly applies for the $\overline{\Sigma}$-point, where
we expect the signature of the subsequently transferred electrons.
 In order to suppress this background, we subtract the energy map
at negative delay from each energy map of the subsequent time steps.
 In addition, these spectra are rebinned by a factor of 8 x 8.
 These difference spectra, which are shown in
Fig.~\ref{fig:excitation}d-h, solely represent the transiently
excited electron distribution.
 In order to account for laser drift during the measurement, we have
repeated the delay scan 10~times with an integration time of 5~min
for each time step resulting in a total acquisition time of
16~hours.

 The difference spectrum in Fig.~\ref{fig:excitation}e shows the
intensity right at the optical excitation at $\overline{K}$.
 It can now be clearly seen that already after one time step of 33~fs
after the optical excitation (Fig. \ref{fig:excitation}f), a
significant population has been built up at the
$\overline{\Sigma}$-point.
 This is a significantly faster transfer time compared to results
deduced from pure optical experiments \cite{Kumar13jap}.
 Remarkably, the population at both points in the conduction band
do not disappear even for delays up to 400~fs
(Fig.~\ref{fig:excitation}h), the longest delay measured in this
experiment.
 We observe an additional broad electron distribution in the lower
left corner of Fig. \ref{fig:excitation}e, which rapidly decays
within the following two time steps.
 We assign this intensity to a direct two-photon transition from the
occupied valence band around the $\overline{\Sigma}$ point without
the population of an intermediate state. Such non-resonant 2PPE is
well known from surface states of metals and its time-dependent
signal has frequently been used to precisely determine the
cross-correlation between pump and probe pulse directly on the
sample in UHV \cite{Hertel96prl,Shumay98prb}. Similarly, we can use
the transient of the region indicated by the black dotted box as
cross-correlation between the visible pump and the HHG probe, which
has a full width at half maximum of 80 fs.

\begin{figure}
    \includegraphics[scale=0.45]{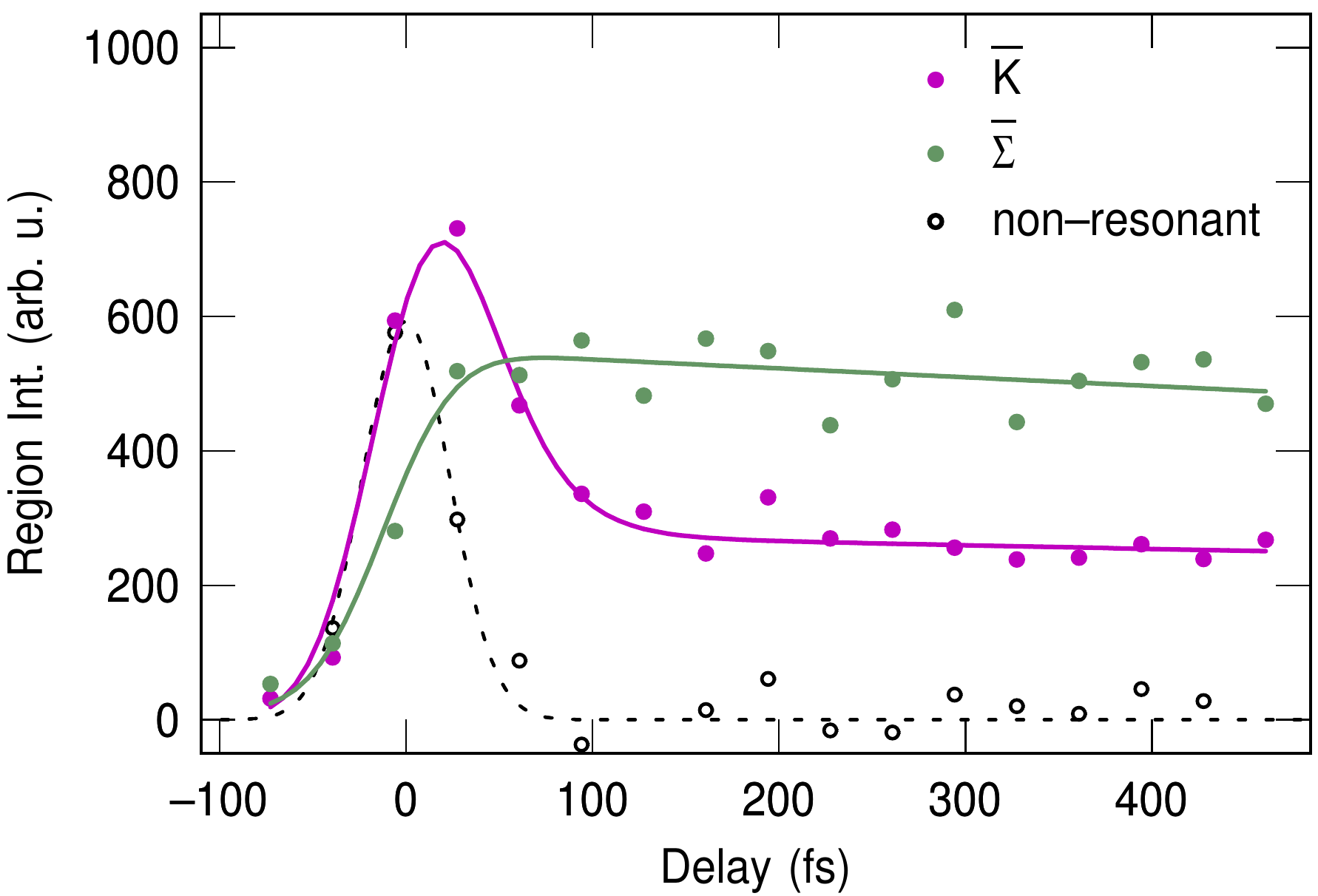}
    \caption{\label{fig:delayscan}
    Temporal evolution of the electron intensities in the two
    regions indicated by green and violet boxes in figure
    \ref{fig:excitation}h. After excitation, the signal at $\protect\overline{K}$ decreases
    almost on the same time scale as it increases while some longer lived signal
    remains. The intensity at $\protect\overline{\Sigma}$ saturates already after one
    time step of 33~fs and then remains constant over
    the measured time interval. The cross-correlation of the laser pulses
    deduced from the the non-resonant signal close
    to $\Sigma$ is indicated in black.}%
\end{figure}

 Fig. \ref{fig:delayscan} shows the temporal evolution of the
electron intensities at $\overline{\Sigma}$ (green) and
$\overline{K}$ (violet) and of the non-resonant 2PPE (black).
 The corresponding integration areas are depicted in
Figs. \ref{fig:excitation}e and h.
 After the optical excitation, the signal at $\overline{K}$ decays first on
an ultrafast time scale and then remains almost constant over the
observed period. It is clear that at least two states have to be
involved in the decay mechanism, one, which decays with a time
constant of $\approx 30$ fs, and one long-lived in the picosecond
range. The signal at $\overline{\Sigma}$ shows an ultrafast increase
and  decays with a time constant indistinguishable from the
long-lived  signal at $\overline{K}$. To understand the transfer
mechanism in detail, further experiments are necessary. A
theoretical work suggests that the transfer time depends on the
excess energy of the excitation \cite{Steinh16arxiv}. With our
excess energy of at least 200 meV, we assume to be on the fast limit
of the transfer. Direct filling of this state at $\overline{\Sigma}$
from the valence band is not possible with our excitation energy due
to the significantly larger gap.

In conclusion, we have presented a direct mapping of the charge
transfer in bulk MoS$_2$ from the $\overline{K}$ to the
$\overline{\Sigma}$ point by time- and angle-resolved two-photon
photoemission with HHG probe pulses.
 We are able to extract a transfer time well below 50~fs, when
exciting the system slightly above the A exciton resonance.
 This is one order of magnitude faster than transfer times reported
previously.
 In both regions, $\overline{\Sigma}$ and $\overline{K}$, a
long-lived population remains in the conduction band with time
constants of several ps.
 Our experimental technique enables many possible applications. The
 aforementioned dependence of the transfer time on the excess energy of the excitation \cite{Steinh162dmat}
 becomes accessible with the tunability of our pump pulses.
 On the other hand, it seems very promising and feasible to observe
the charge transfer after spin-selective excitation of the distinct
$\overline{K}$ and $\overline{K}'$ points using circularly polarized
pump pulses even in bulk samples \cite{Riley14natphys,
Zhang14natphys}.
 With a 3D analyzer, it is even possible to access different high
symmetry points at the same time without moving the sample and
thereby observe the charge distribution in the non-equivalent
$\overline{\Sigma}$ and $\overline{\Sigma}'$ points.
 By exciting the system on and above resonance, signatures of the
exciton should be observable in the photoelectron distribution which
would enable to address many open questions concerning the exciton
dynamics in these materials. Particulary in the case of van der
Waals heterostructures different relaxation pathways which are
indistinguishable in real space could be traced in momentum space.

We gratefully acknowledge funding by the Deutsche
Forschungsgemeinschaft through the SFB 1083. We thank C. Heyl for
valuable discussion concerning the improvement of the high-harmonic
generation.
%

\end{document}